\documentclass[a4paper,11pt]{article}
\pdfoutput=1 

\usepackage{jheppub} 

\usepackage[T1]{fontenc} 
\usepackage{arydshln}

\usepackage{graphicx}
\usepackage{tikz}
\tikzset{>=latex}
\usetikzlibrary{trees}
\usetikzlibrary{decorations.pathmorphing}
\usetikzlibrary{decorations.markings}
\usetikzlibrary{decorations.pathreplacing}
\tikzset{
    gluon/.style={decorate, draw=black,
        decoration={coil,amplitude=3pt, segment length=4pt,aspect=0.7}} 
}
\tikzset{
    gluon2/.style={decorate, draw=black,
        decoration={coil,amplitude=2.5pt, segment length=2.5pt,aspect=0.7}} 
}
\tikzset{
    gluon3/.style={decorate, draw=black,
        decoration={coil,amplitude=2.pt, segment length=2.pt,aspect=0.7}} 
}

\newcommand{\be}{\begin{eqnarray}}
\newcommand{\ee}{\end{eqnarray}}
\newcommand{\sig}{\sigma}
\newcommand{\ma}{\mathcal{A}}
\newcommand{\mc}{\mathcal{C}}
\def\ol#1{\overline{#1}}

\title{Proof of a new colour decomposition for QCD amplitudes}

\author[a,b]{Tom Melia}

\affiliation[a]{Department of Physics, University of California, Berkeley, California 94720, USA}
\affiliation[b]{Theoretical Physics Group, Lawrence Berkeley National Laboratory, Berkeley, California 94720, USA}

\emailAdd{tmelia@lbl.gov}

\preprint{UCB-PTH-15/06}

\abstract{Recently, Johansson and Ochirov conjectured the form of a  new colour decomposition for QCD tree-level amplitudes. This note provides a proof of that conjecture. The proof is based on `Mario World' Feynman diagrams, which exhibit the hierarchical Dyck structure previously found to be very useful when dealing with multi-quark amplitudes.
\begin{figure}[b]
\begin{center}
\includegraphics{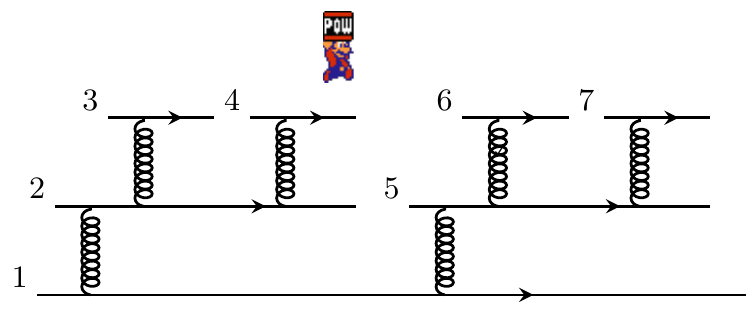}
\end{center}
\end{figure}
}

\notoc 
\begin{document} 
\maketitle
\flushbottom
\newpage

\section{Introduction and discussion}

A colour decomposition of a gauge theory scattering amplitude, $\ma$,  takes the schematic form
\be
\ma =\sum_i C_i\, A_i\,,
\label{eq:coldeca}
\ee
where $A_i$ are kinematic primitive amplitudes and $C_i$ are colour factors. Let $\ma_{n,k}$ be a tree-level QCD\footnote{In this note I will use the language of QCD---gluons and quarks---but gluon can be taken to mean the gauge field associated with any Lie group, and quark can be taken to mean matter transforming in any representation of the group.}  amplitude with $k$ quark-antiquark pairs and $n-2k$ gluons. A colour decomposition for all-gluon trees is \cite{DelDuca:1999rs} 
\be
\ma_{n,0}&=&  \sum_{\sig\in S_{n-2}}( F^{\sig_1}\ldots F^{\sig_{n-2}})^{1}_{~n} \, A(1,\sig, n) \,,
\label{eq:glucol}
\ee
where $F^{abc}$ are the structure constants. The primitives, $A(1,\sig,n)$, are gauge invariant, and have a fixed cyclic order of external legs, to which the permutation $\sig$ corresponds. Eq.~\eqref{eq:glucol} is an example of what I will refer to as a {\it proper} colour decomposition in that no further group theory induced relations, known as the Kleiss-Kuijf (KK) relations \cite{Kleiss:1988ne}, exist between the primitives used in the colour decomposition. That is, eq.~\eqref{eq:coldeca} is a proper colour decomposition when the sum is over a minimal set of KK-independent primitive amplitudes; in the case of gluon trees, this set is of size $(n-2)!$. Colour decompositions are not necessarily proper (the well-known trace-baced decomposition for gluons \cite{Berends:1987cv,Mangano:1987xk} is not), but  can always be made so through application of the KK relations to re-express  primitives in terms of those in a chosen KK-independent basis. A proper colour decomposition for amplitudes with one quark-antiquark pair is \cite{Mangano:1988kk},
\be
\ma_{n,1} &=& \sum_{\sig\in S_{n-2}} \{q|T^{\sig_1}\ldots T^{\sig_{n-2}} | q\}  \,A(q,\sig,\ol{q}) \,,
\label{eq:quarkcol}
\ee
where  $T^a$ are matrices in the fundamental representation of  $SU(3)$, $\{q| T^a |q\} = (T^a)^{q}_{~\ol{q}}$ in more conventional notation, and again the basis of primitives is of size $(n-2)!$.

When more quark lines are present, the concept of defining a primitive amplitude by demanding a cyclic ordering of external legs {\it a priori} was first considered in \cite{Bern:1994fz}. The papers \cite{Melia:2013bta,Melia:2013epa} explored the basic properties of the general QCD primitive tree with $k$ quark-antiquark pairs and $n-2k$ gluons. In particular, it was shown how matter and flavour modify the group-theoretic relations between the primitives by introducing additional structure based around Dyck words. When all $k$ quark pairs have distinct flavours, a KK-independent basis of primitives is of size $(n-2)!/k!$.

Proper colour decompositions for amplitudes with two and three quark lines were previously presented in the proceedings \cite{Melia:2014oza}.  Recently, Johansson and Ochirov (JO) used the Dyck basis to make a conjecture as to the form of a proper colour decomposition for a general QCD tree \cite{Johansson:2015oia}. Given the relative complexity of the expressions obtained via known methods of constructing QCD colour decompositions for specific cases \cite{Ellis:2011cr,Ita:2011ar,Badger:2012pg,Reuschle:2013qna}\footnote{These methods do not in general yield a proper colour decomposition; they may contain primitives which are zero, and redundancy relations between primitives may remain.}, the JO conjecture is quite remarkable.  This note is a proof of this conjecture.

The proof I will present  is diagrammatic; it is based ultimately on Feynman diagrams. This highlights the fact that the decomposition is gauge group and matter representation independent, as the primitive colour factors are directly related to Feynman diagram colour factors. A particular class of Feynman diagrams---the `Mario Worlds'---turn out to be very useful for the proof.

The `Mario World' diagrams display the Dyck structure uncovered in \cite{Melia:2013bta,Melia:2013epa}, which has also been important in understanding further properties of  general QCD primitives: very recently they have been shown to obey Bern-Carrasco-Johansson relations  \cite{Bern:2008qj,Johansson:2014zca,Johansson:2015oia,delaCruz:2015dpa}  and to have a Cachazo-He-Yuang representation \cite{delaCruz:2015raa}. I expect the `Mario World' diagram construction to be useful for any diagrammatic study of general colour decompositions beyond tree level, although an understanding of the one-loop QCD primitive basis is still currently lacking.
 
The remainder (section~\ref{sec:two}) of this note is the proof of the JO conjectured colour decomposition, which is stated in eq.~\eqref{eq:jo}.

\section{Proof of the new colour decomposition for QCD trees}
\label{sec:two}
The outline of the proof is as follows. In section~\ref{2p1} I present the JO conjecture. In section~\ref{2p2} I show that the colour factors $C$ satisfy a recursion relation which will form the basis of a proof by induction. In section~\ref{2p3}, I describe the set-up of a system of linear equations which relates primitive colour factors to Feynman diagram colour factors. In  section~\ref{2p4} I introduce `Mario World' Feynman diagrams, and the linear system of equations they belong to. In section~\ref{2p5} I use this system to show that the recursion relation in section~\ref{2p2} is satisfied.

\subsection{The JO conjecture}
\label{2p1}

The conjectured colour decomposition is for $n$-point tree-level QCD amplitudes with $k$ distinct flavour, possibly massive, quark-antiquark pairs and $n-2k$ gluons. The basis taken for the kinematic, primitive amplitudes is that of \cite{Melia:2013bta,Melia:2013epa}  based around Dyck words. A Dyck word of length $2k$ can be defined as a composition of $k$ pairs of parentheses `$($' and `$)$', such that they are closed correctly. Identified pairs of parentheses are those which close with each other. The level of an identified pair is the level of nestedness they reside at.\footnote{{\it i.e.} the number of `$($' minus the number of  `$)$' to the left of the opening parenthesis of the pair.} The set of primitive amplitudes for the basis is written as
\be
\{ A(1,\sigma,\ol{1})  \,\, | \,\, \sigma \in \text{Dyck}_{k-1} \} \,,
\label{ref:basis}
\ee
where Dyck$_{k-1}$ means all possible permutations of quark and gluon labels where the quark-antiquark (unbarred and barred labels, respectively) flavour pairs form an identified pair of parentheses in a Dyck word, with the quark label being the opening parenthesis.\footnote{Whether to assign quark or antiquark to the opening parenthesis can be decided independently for each flavour pair in each Dyck word, which was a main result of \cite{Melia:2013bta,Melia:2013epa}. The orientation of the quark lines chosen in this paper is the convention of \cite{Melia:2013epa}.}  For example, for $n=6$, $k=3$, we have Dyck$_{k-1}=\{ 2 3\ol{3} \ol{2} ,  3 2\ol{2} \ol{3} ,   2 \ol{2} 3 \ol{3} , 3 \ol{3} 2 \ol{2 }\}$; for $n=7$, $k=3$,  we have Dyck$_{k-1}=\{g 2 3\ol{3} \ol{2} ,2g 3\ol{3} \ol{2} ,2 3g\ol{3} \ol{2} ,2 3\ol{3}g \ol{2} ,2 3\ol{3} \ol{2}g ,\ldots  \}$, where $g$ is the gluon label, and where `$\ldots$' stands for gluon insertions around the  three other Dyck words.

\begin{figure}
\centering
\begin{tikzpicture}

\begin{scope}[thick,decoration={
    markings,
    mark=at position 0.8 with {\arrow{stealth}}}
    ]

\draw[-,postaction={decorate}] (0,0)--(1,0); 
\draw[-,postaction={decorate}] (1,0)--(2,0); 
\draw[-,postaction={decorate}] (2,0)--(3,0); 

\draw[gluon] (0.4,0)--(0.4,1); 
\draw[gluon] (1.4,0)--(1.4,1); 
\draw[gluon] (2.4,0)--(2.4,1); 

\filldraw[fill=white] (0.4,0) circle (0.1cm);
\filldraw[fill=white] (1.4,0) circle (0.1cm);
\filldraw[fill=white] (2.4,0) circle (0.1cm);

\draw[-,postaction={decorate}] (0,1)--(1,1); 
\draw[-,postaction={decorate}] (1,1)--(2,1); 
\draw[-,postaction={decorate}] (2,1)--(3,1); 

\draw[-,postaction={decorate}] (1.1,2)--(1.9,2); 
\draw[-,postaction={decorate}] (2.1,2)--(2.9,2); 

\draw[gluon] (1.4,1)--(1.4,2); 
\draw[gluon] (2.4,1)--(2.4,2); 

\filldraw[fill=white] (1.4,1) circle (0.1cm);
\filldraw[fill=white] (2.4,1) circle (0.1cm);

\node at (-0.1,0.2) {\small \( 1  \)};
\node at (-0.1,1.2) {\small \( 2  \)};
\node at (1,2.2) {\small \( 3  \)};
\node at (2.1,2.2) {\small \( 4  \)};

\node at (1.5,-0.7) { \( C_{1\,2\,3\,\ol{3}\,4\,\ol{4}\,\ol{2}\,\ol{1} } \)};

\end{scope}

\end{tikzpicture}
\hspace{1.3cm}
\begin{tikzpicture}

\begin{scope}[thick,decoration={
    markings,
    mark=at position 0.8 with {\arrow{stealth}}}
    ] 
    
\draw[-,postaction={decorate}] (0,0)--(1,0); 
\draw[-,postaction={decorate}] (1,0)--(2,0); 
\draw[-,postaction={decorate}] (2,0)--(3,0); 
\draw[-] (3,0)--(3.5,0); 
\draw[-,postaction={decorate}] (3.5,0)--(4.5,0); 
\draw[-,postaction={decorate}] (4.5,0)--(5.5,0); 

\draw[gluon] (0.4,0)--(0.4,1); 
\draw[gluon] (1.4,0)--(1.4,1); 
\draw[gluon] (2.4,0)--(2.4,1); 
\draw[gluon] (3.9,0)--(3.9,1); 
\draw[gluon] (4.9,0)--(4.9,1); 

\filldraw[fill=white] (0.4,0) circle (0.1cm);
\filldraw[fill=white] (1.4,0) circle (0.1cm);
\filldraw[fill=white] (2.4,0) circle (0.1cm);
\filldraw[fill=white] (3.9,0) circle (0.1cm);
\filldraw[fill=white] (4.9,0) circle (0.1cm);

\draw[-,postaction={decorate}] (0,1)--(1,1); 
\draw[-,postaction={decorate}] (1,1)--(2,1); 
\draw[-,postaction={decorate}] (2,1)--(3,1); 
\draw[-,postaction={decorate}] (3.5,1)--(4.5,1); 
\draw[-,postaction={decorate}] (4.5,1)--(5.5,1); 

\draw[-,postaction={decorate}] (1.1,2)--(2,2); 
\draw[-,postaction={decorate}] (2,2)--(3,2); 

\draw[gluon] (1.4,1)--(1.4,2); 
\draw[gluon] (2.4,1)--(2.4,2); 
\draw[gluon] (4.9,1)--(4.9,2); 

\filldraw[fill=white] (1.4,1) circle (0.1cm);
\filldraw[fill=white] (2.4,1) circle (0.1cm);
\filldraw[fill=white] (4.9,1) circle (0.1cm);

\draw[gluon] (2.4,2)--(2.4,3); 
\filldraw[fill=white] (2.4,2) circle (0.1cm);

\draw[-,postaction={decorate}] (2.1,3)--(2.9,3); 

\node at (-0.1,0.2) {\small \( 1  \)};
\node at (-0.1,1.2) {\small \( 2  \)};
\node at (1,2.2) {\small \( 3  \)};
\node at (1.9,3.) {\small \( 4  \)};
\node at (3.4,1.2) {\small \( 5  \)};
\node at (5.,2.2) {\small \( g  \)};

\node at (2.7,-0.7) { \( C_{1\,2\,3\,4\,\ol{4}\,\ol{3}\,\ol{2}\,5\,g\,\ol{5}\,\ol{1} } \)};

\end{scope}

\end{tikzpicture}
\caption{Johansson and Ochirov's diagrammatic notation for primitive  colour factors. The colour factor on the left consists of four terms upon expanding out the $\Xi$ operators (see text for explicit expressions). The one on the right consists of twelve such terms. }
\label{fig:colfac}
\end{figure}
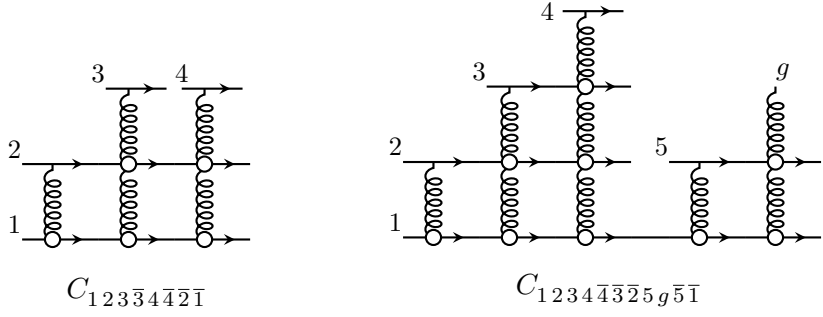

As the basis~\eqref{ref:basis} was shown in \cite{Melia:2013bta,Melia:2013epa}  to consist of independent primitives under KK relations, then
\be
\ma_{n,k} = \sum_{\sigma\in \text{Dyck}_{k-1}} C_{1 \, \sigma \, \ol{1}} \, A(1,\sigma,\ol{1}) \,,
\ee
is a proper colour decomposition. The JO conjecture is that the colour factor has a direct relationship with the permutation $\sigma$ (there is a slight difference in convention between the basis chosen in \cite{Johansson:2015oia} and that in this note):
\be
C_{1 \, \sigma \, \ol{1}} = \{ 1| \sigma | 1\} \bigg|_{ \begin{subarray}{l}  q\to \{q | T^b \, \otimes\, \Xi^b_{l} \\ g\to \Xi^{a_g}_{l}  \\ \ol{q}\to | q\}  \end{subarray} } \,.
\label{eq:jo}
\ee
The above notation means the following. To build up the colour factor, one reads the permutation $\sig$ from left to right. Whenever a quark label is encountered, add to the colour factor the combination $ \{q | T^b \, \otimes\, \Xi^b_{l} $, where
\be
\Xi^b_{l} = \sum_{s=1}^{l} \underbrace{1\,\otimes \,\ldots \,\otimes 1\, \otimes  \, \overbrace{T^b\, \otimes\, 1\, \otimes \,\ldots \,\otimes \, 1}^s }_{l} \,,
\label{eq:xidef}
\ee
and $l$ is the level of nestedness of the quark  $q$ in the Dyck word. Whenever a gluon is encountered, add to the colour factor $\Xi^{a_g}_l$, where $l$ is the level of nestedness of the gluon. Whenever an antiquark label is encountered, add to the colour factor $|q\}$. 
The tensors in the tensor product in eq.~\eqref{eq:xidef} act on quark lines at different levels; the leftmost tensor at level $l-1$, and the rightmost at level $0$. Johansson and Ochirov represent the operator $\Xi$ diagrammatically
\be\begin{tikzpicture}

\begin{scope}[thick,decoration={
    markings,
    mark=at position 0.8 with {\arrow{stealth}}}
    ]     
\node  at (-0.4,0) { \( \Xi^{a}_l \)};
\node  at (0.2,0) { \(  =  \)};

\draw[-,postaction={decorate}] (0.8,-1)--(1.8,-1); 
\draw[-,postaction={decorate}] (0.8,-0.3333)--(1.8,-0.3333); 
\draw[-,postaction={decorate}] (0.8, 0.3333)--(1.8, 0.3333); 
\draw[-,postaction={decorate}] (0.8,1)--(1.8,1); 

\draw[gluon] (1.2,-1)--(1.2,-0.3333); 
\draw[gluon] (1.2,-0.3333)--(1.2,0.3333); 
\draw[gluon] (1.2,0.3333)--(1.2,1); 
\draw[gluon] (1.2,1)--(1.2,1.6666); 

\node  at (1.2,1.8533) { \(  a  \)};

\filldraw[fill=black] (1.5,0.5) circle (0.01cm);
\filldraw[fill=black] (1.5,0.65) circle (0.01cm);
\filldraw[fill=black] (1.5,0.8) circle (0.01cm);

\filldraw[fill=white] (1.2,-1) circle (0.1cm);
\filldraw[fill=white] (1.2,-0.3333) circle (0.1cm);
\filldraw[fill=white] (1.2, 0.3333) circle (0.1cm);
\filldraw[fill=white] (1.2, 1) circle (0.1cm);

\draw [decorate,decoration={brace,amplitude=5pt,mirror,raise=2pt},yshift=0pt]
(1.8,-1) -- (1.8,1) node [black,midway,xshift=0.4cm] {\footnotesize
$l$};
\node  at (2.8,0) { \(  =  \)};


\draw[-,postaction={decorate}] (2.5+0.8,-1)--(2.5+1.8,-1); 
\draw[-,postaction={decorate}] (2.5+0.8,-0.3333)--(2.5+1.8,-0.3333); 
\draw[-,postaction={decorate}] (2.5+0.8, 0.3333)--(2.5+1.8, 0.3333); 
\draw[-,postaction={decorate}] (2.5+0.8,1)--(2.5+1.8,1); 

\filldraw[fill=black] (2.5+1.5,0.5) circle (0.01cm);
\filldraw[fill=black] (2.5+1.5,0.65) circle (0.01cm);
\filldraw[fill=black] (2.5+1.5,0.8) circle (0.01cm);

\draw[fill=white,white] (2.5+1.05,-0.8) rectangle (2.5+1.35,1.6666);
\draw[gluon] (2.5+1.2,-1)--(2.5+1.2,1.6666); 

\node  at (2.5+2.3,0) { \(  +  \)};


\draw[-,postaction={decorate}] (4.5+0.8,-1)--(4.5+1.8,-1); 
\draw[-,postaction={decorate}] (4.5+0.8,-0.3333)--(4.5+1.8,-0.3333); 
\draw[-,postaction={decorate}] (4.5+0.8, 0.3333)--(4.5+1.8, 0.3333); 
\draw[-,postaction={decorate}] (4.5+0.8,1)--(4.5+1.8,1); 

\filldraw[fill=black] (4.5+1.5,0.5) circle (0.01cm);
\filldraw[fill=black] (4.5+1.5,0.65) circle (0.01cm);
\filldraw[fill=black] (4.5+1.5,0.8) circle (0.01cm);

\draw[fill=white,white] (4.5+1.05,-0.2) rectangle (4.5+1.35,1.6666);
\filldraw[fill=white,gluon] (4.5+1.2,-0.3333)--(4.5+1.2,1.6666); 

\node  at (4.5+2.4,0) { \(  +  \)};
\node  at (5.1+2.4,0) { \(  \ldots  \)};
\node  at (5.7+2.4,0) { \(  +  \)};


\draw[-,postaction={decorate}] (7.8+0.8,-1)--(7.8+1.8,-1); 
\draw[-,postaction={decorate}] (7.8+0.8,-0.3333)--(7.8+1.8,-0.3333); 
\draw[-,postaction={decorate}] (7.8+0.8, 0.3333)--(7.8+1.8, 0.3333); 
\draw[-,postaction={decorate}] (7.8+0.8,1)--(7.8+1.8,1); 

\filldraw[fill=black] (7.8+1.5,0.5) circle (0.01cm);
\filldraw[fill=black] (7.8+1.5,0.65) circle (0.01cm);
\filldraw[fill=black] (7.8+1.5,0.8) circle (0.01cm);

\filldraw[fill=white,gluon] (7.8+1.2,1.)--(7.8+1.2,1.6666); 

\end{scope}

\end{tikzpicture} 
\label{eq:xidia}\ee
This diagrammatic notation extends to provide a useful shorthand for the entire colour factor---see fig.~\ref{fig:colfac} for two examples. The number of terms that result from expanding out the sum in the $\Xi$ operators is $\prod_q l_q\prod_g l_g $, where $q$ runs over all quark pairs other than flavour pair $1$, with $l_i$ being the level of the quark pair; $g$ runs over all gluons, with $l_g$ being the level of the gluon. As an example, the  colour factor on the left in fig.~\ref{fig:colfac} has four terms when expanded in this way: 
\be
&\{1| T^a |1\} \{2| T^a T^b T^c|2\} \{3| T^b |3\} \{4| T^c |4\} ~~~~\{1| T^a  T^c|1\} \{2| T^a T^b|2\} \{3| T^b |3\} \{4| T^c |4\}\nonumber\\
&\{1| T^a T^b |1\} \{2| T^a T^c|2\} \{3| T^b |3\} \{4| T^c |4\} ~~~~\{1| T^a T^b T^c |1\} \{2| T^a|2\} \{3| T^b |3\} \{4| T^c |4\} \nonumber.
\ee

I end this subsection with two further points. Firstly, I define a quantity that will play an important role in the following:  the sum over the levels of all quarks and gluons, $l_{\text{tot}}=\sum_{i=q,g}l_i$.  This will serve as a hierarchical parameter to set up a recursion.

Secondly, note for the purpose of discerning the colour decomposition, gluons can be considered as just a special case of a quark line: one that always appears with consecutive quark-antiquark labels ({\it i.e.} no further quark lines nest above it). 
For example, the quark lines labelled 3 and 4 on the left diagram in fig.~\ref{fig:colfac} could represent gluons with a replacement of the colour factor $\{3| T^{a} |3\}\to \delta^{a \,a_{g_3}}$ and  $\{4| T^{a} |4\}\to \delta^{a \,a_{g_4}}$ (an interpretation of the $gq\ol{q}$ vertex as a gluon polarisation vector is required in the accompanying kinematic part). For the diagram on the right of fig.~\ref{fig:colfac}, only the quark line labelled 4 could represent a gluon. It is  sufficient to prove the colour decomposition eq.~\eqref{eq:jo} for the all-quark case, with the understanding that such replacements can be made to recover the case with gluons.

\subsection{Recursion relation}
\label{2p2}

The colour factors defined above satisfy a recursion relation. I define a {\it leading quark line} as a quark line $a$ at a level $l_a>1$ that appears in the permutation as $\sig=(\ldots a,\ol{a},\ol{b}\ldots)$, where $b$ is another quark line at level $l_a-1>0$.  There is always one such quark line except in the case where all quark lines are at level $l=1$. Quark line 4  is a leading quark line in both diagrams in fig.~\ref{fig:colfac}.

\begin{figure}
\centering
\makebox[\textwidth][c]{
\input{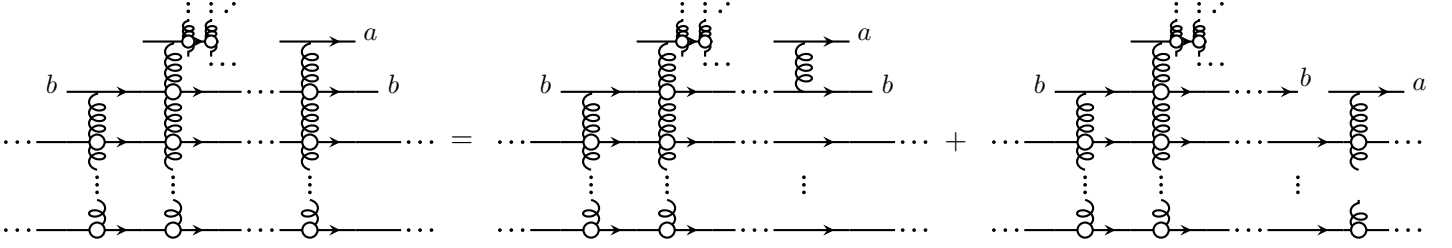}
}
\caption{Diagrammatic representation of the recursion relation, eq.~\eqref{eq:rec1}, for primitive colour factors.}
\label{fig:rec}
\end{figure}

The recursion follows from eqs.~\eqref{eq:xidef},~\eqref{eq:xidia}, splitting up the last term (where the gluon connects to the quark line of highest level---$s=l$ in eq.~\eqref{eq:xidef}), and the rest of the terms:
\be
&~&\!\!\!\!\!\!\!\!\!\!\!\!\!\!\!\!\!\!\!\!\!\!\!\{1| \ldots \{b|\ldots \{ a| \,T^c \otimes \Xi^{c}_{l_a}\, | a\}|b\} \ldots |1\} \nonumber \\
 &~&\!\!\!\! \!\!\!\!\!\!\!\!= \{1| \ldots \{b|\ldots \{ a| \,T^c \, | a\} T^c|b\} \ldots |1\}  \,\,+\,\, \{1| \ldots \{b|\ldots|b\}  \{ a| \,T^c \otimes \Xi^{c}_{l_a-1}\, | a\}\ldots |1\} \label{eq:rec1}\\
 &~&\!\!\!\! \!\!\!\!\!\!\!\!= \{1| \ldots \{b|\ldots |ab\} \ldots |1\} \,\,+ \,\,\{1| \ldots \{b|\ldots|b\}  \{ a| \,T^c \otimes \Xi^{c}_{l_a-1}\, | a\}\ldots |1\}   \,,
\label{eq:rec2}
\ee
which in terms of primitive colour factors is,
\be
C_{..a  \,\ol{a} \, \ol{b}..} = C_{  .. \ol{ab}..} + C_{.. \ol{b} \, a\, \ol{a} ..} \,.
\label{eq:rec3}
\ee
The recursion is diagrammatically represented in fig.~\ref{fig:rec}.

The first term on the rhs of eqs.~\eqref{eq:rec2},~\eqref{eq:rec3} is written in such a way to emphasise that this is a colour factor for an amplitude where quark line $a$ has been removed, but where the colour factor associated with the antiquark $b$ is substituted with a `colour polarisation' to include the quark line $a$'s colour factor {\it i.e.} $|b\}\to \{a| T^c |a\}T^c |b\}\equiv |ab\}$. (In the colour factor label, the three labels $a \,\ol{a}\,\ol{b}$ become one label, $\ol{ab}$.) In the kinematic part, the polarisation of the antiquark $\ol{b}$ is suitably modified to incorporate the effect of quark line $a$. Both terms on the rhs of eq.~\eqref{eq:rec3} now have lower $l_{\text{tot}}$: the first term has $l_{\text{tot}}-l_a$ as quark line $a$ has been removed; the second term has $l_{\text{tot}}-1$ as the quark line $a$ is now nested at one level lower. This recursion relation allows for a proof by induction using $l_{\text{tot}}$. 

\subsection{Feynman diagram system}
\label{2p3}

Now I set up a system of linear equations relating primitive colour factors to Feynman diagram colour factors.
I write the amplitude in two ways: as a sum over Feynman diagrams, where I separate out the colour part $\mc_j$ from the kinematic part $D_j$ (both given by standard Feynman rules) of the diagram, and as a sum over primitive amplitudes $A_i$ multiplied by their colour factors $C_i$,
\be
\ma= \sum_j \mc_{j} \, D_j = \sum_{i} C_{i} \, A_i  \,.
\ee
The primitive amplitudes can be written in terms of the individual kinematic parts $D_j$ modulo the colour-ordered Feynman rules: $A_i=\sum_j M_{ji}D_j$, where $M_{ji}$ is $\pm 1$ if the diagram contributes to the primitive (the sign depends on the relative ordering of the legs at each vertex in the diagram---exchanging two legs induces a negative sign under the colour-ordered rules), or else $0$. This implies the following relation between the primitive colour factors and the Feynman colour factors,
\be
  \mc_{j}=\sum_{i}  M_{ji} \,C_{i}  \,.
\label{eq:matrel}
\ee
Knowing the Feynman colour factors, $\mc_j$ and the rectangular matrix $M_{ji}$, one can pseudo-invert the system to obtain the primitive colour factors, $C_i$. (This system, of course, depends on the choice of primitive basis.) It is useful to rewrite eq.~\eqref{eq:matrel} in the form
\be
\mc_j =\sum_{\sig\in\text{flips}_j} (-1)^\sig\, C_\sig  \,,
\label{eq:sigrel}
\ee
where $\{\text{flips}_j\}$ are all permutations  {\it in the basis}  corresponding to possible ways of drawing the Feynman diagram associated with $\mc_j$ in a planar fashion. For an $n=4$, $k=1$ example, the four planar drawings corresponding to all possible flips of the two vertices for the  diagram $q\ol{q}\to g\to g_1g_2$ are,
\begin{center}
\begin{tikzpicture}[scale=1.0]

\begin{scope}[thick,decoration={
    markings,
    mark=at position 0.8 with {\arrow{stealth}}}
    ]

\draw[-,postaction={decorate}] (-0.3+0,0)--(-0.3+2,0); 

\draw[gluon] (-0.3+1.,0.0)--(-0.3+1.,1.); 
\draw[gluon] (-0.3+0.,1.)--(-0.3+2.,1.0); 

\node[black] at (-0.3+-0.2,0.0) {\small \( q  \)};
\node[black] at (-0.3+-.2,1.) {\small \( g_1  \)};
\node[black] at (-0.3+2.2,0.) {\small \( \ol{q}  \)};
\node[black] at (-0.3+2.2,1.) {\small \( g_2  \)};

\draw[-,postaction={decorate}] (3.0+ 0,0)--(3.0+ 2,0); 

\draw[gluon] (3.0+ 1.,0.0)--(3.0+ 1.,1.); 
\draw[gluon] (3.0+ 0.,1.)--(3.0+ 2.,1.0); 

\node[black] at (3.0+ -0.2,0.0) {\small \( q  \)};
\node[black] at (3.0+ -.2,1.) {\small \( g_2  \)};
\node[black] at (3.0+ 2.2,0.) {\small \( \ol{q}  \)};
\node[black] at (3.0+ 2.2,1.) {\small \( g_1  \)};

\draw[-,postaction={decorate}] (6.5+ 0,0)--(6.5+ 0.0,1.); 

\draw[gluon] (6.5+ 0.,0.5)--(6.0+ 2., 0.5); 
\draw[gluon] (6.0+ 2.,0.)--(6.0+ 2.,1.0); 

\node[black] at (6.5+ -0.2,0.0) {\small \( q  \)};
\node[black] at (6.5+ -.2,1.) {\small \( \ol{q}  \)};
\node[black] at (6.0+ 2.2,0.) {\small \( g_2  \)};
\node[black] at (6.0+ 2.2,1.) {\small \( g_1  \)};

\draw[-,postaction={decorate}] (9.5+ 0,0)--(9.5+ 0.0,1.); 

\draw[gluon] (9.5+ 0.,0.5)--(9.0+ 2., 0.5); 
\draw[gluon] (9.0+ 2.,0.)--(9.0+ 2.,1.0); 

\node[black] at (9.5+ -0.2,0.0) {\small \( q  \)};
\node[black] at (9.5+ -.2,1.) {\small \( \ol{q}  \)};
\node[black] at (9.0+ 2.2,0.) {\small \( g_1  \)};
\node[black] at (9.0+ 2.2,1.) {\small \( g_2  \)};

\end{scope}

\end{tikzpicture} \,,
\end{center}
where the relative signs of the permutation are $(+,-,-,+)$. However, if the chosen basis for the primitives was $\ma(q,\sig,\ol{q})$, then only the first two diagrams correspond to permutations within the basis. In this case then $\{\text{flips}\} = \{qg_1g_2\ol{q},\, qg_2g_1\ol{q}\}$, and eq.~\eqref{eq:sigrel} reads $\mc_{q\ol{q}\to g\to g_1g_2}=C_{qg_1g_2\ol{q}}-C_{qg_2g_1\ol{q}}$. In the following we will only encounter diagrams with trivalent vertices, and so the number of possible flips is $2^{\#\text{vertices}}$ (although, in general, and as in the example above, not all flips will correspond to permutations within the chosen basis).

As an example, let us see how this works for an amplitude with one quark line and $n-2$ gluons. Here we can take as a basis the $(n-2)!$ primitives of the form $A(q,\sig_1,\ldots,\sig_{n-2},\ol{q})$, where $\sig$ gives the permutation of the $(n-2)$ gluon labels. It turns out that there is a subset of Feynman diagrams which are particularly useful for our purpose of inverting the system of equations in eq.~\eqref{eq:matrel}; these are `comb' diagrams which have all the gluons individually attached to the quark line:
\begin{center}
\begin{tikzpicture}[scale=1.4]

\begin{scope}[thick,decoration={
    markings,
    mark=at position 0.7 with {\arrow{stealth}}}
    ]

\draw[-,postaction={decorate}] (0,0)--(4,0); 

\draw[gluon2] (0.3,0.0)--(0.3,0.5); 
\draw[gluon2] (0.8,0.0)--(0.8,0.5); 
\draw[gluon2] (1.3,0.0)--(1.3,0.5); 
\draw[gluon2] (1.8,0.0)--(1.8,0.5); 
\filldraw[fill=black] (2.4,0.25) circle (0.01cm);
\filldraw[fill=black] (2.7,0.25) circle (0.01cm);
\filldraw[fill=black] (3,0.25) circle (0.01cm);

\draw[gluon2] (3.6,0.0)--(3.6,0.5); 

\node[black] at (4+0.2,0.0) {\small \( \ol{q}  \)};
\node[black] at (-0.2,0.0) {\small \( q  \)};
\node[black] at (0.3,0.7) {\small \( g_{\sigma_1}  \)};
\node[black] at (0.8,0.7) {\small \( g_{\sigma_2}  \)};
\node[black] at (1.3,0.7) {\small \( g_{\sigma_3}  \)};
\node[black] at (1.8,0.7) {\small \( g_{\sigma_4}  \)};
\node[black] at (3.6,0.7) {\small \( g_{\sigma_{n-2}}  \)};

\end{scope}

\end{tikzpicture} \,.
\end{center}
Each of these Feynman diagrams only contributes to one primitive amplitude---a flip of any vertex in the comb gives a planar diagram which does not contribute to a primitive in the chosen basis. (In \cite{Melia:2014oza} I termed such a diagram a unique Feynman diagram, or UFD; with a careful choice of basis, UFDs can be found for all four quark, and most six quark amplitudes.) Note that this argument relies on the fact that it is only the colour factors of primitives to which the Feynman diagram contributes that appear on the rhs of eq.~\eqref{eq:sigrel}. Writing out the matrices in eq.~\eqref{eq:matrel} explicitly, we have found a subset of Feynman diagrams where the relation between primitive colour factor and Feynman diagram colour factor is trivial:
\be
 \left(   
 \begin{array}{c }
 \\
 \\
\hdashline
\mc^{\text{comb}}_1 \\
\mc^{\text{comb}}_{2} \\
.\\
.\\
.\\
\mc^{\text{comb}}_{(n-2)!} \\
\hdashline
\\
\\
\end{array}
\right) 
=
\left(   
\begin{array}{c c c c cc }
 &  & &  &   \\
 &  & &  &   \\
\hdashline
1 & 0 & 0 & . & . & 0 \\
0 & 1 &  0 & . & . & 0 \\
 0& 0&1 & &   & . \\
 &  &  & .& &  . \\
. &  &  & & . &  0 \\
0 & 0 &.&  .&  0 & 1 \\
\hdashline
 &  & &  &   \\
 &  & &  &   \
\end{array}
\right)
 \cdot
 \left(   
 \begin{array}{c }
C_{1} \\
C_{2} \\
.\\
.\\
.\\
C_{(n-2)!} \\
\end{array}
\right) \,.
\ee
The primitive colour factor is exactly the Feynman diagram colour factor of the corresponding comb UFD---$\{q|T^{\sig_1}\ldots T^{\sig_{n-2}} |q\}$---which reproduces the known colour decomposition eq.~\eqref{eq:quarkcol}. An identical argument with the quark line in the comb replaced by a gluon line reproduces eq.~\eqref{eq:glucol} for gluon trees.

\subsection{`Mario World' diagrams}
\label{2p4}

\begin{figure}
\centering
\begin{tikzpicture}[scale=1.4]

\begin{scope}[thick,decoration={
    markings,
    mark=at position 0.7 with {\arrow{stealth}}}
    ]

\draw[-,postaction={decorate}] (0,0)--(4,0); 

\draw[-,postaction={decorate}] (0.1,0.5)--(1.8,0.5); 
\draw[-,postaction={decorate}] (2.1,0.5)--(3.8,0.5); 

\draw[-,postaction={decorate}] (0.4,1.0)--(1.0,1.0); 
\draw[-,postaction={decorate}] (1.2,1.0)--(1.8,1.0); 

\draw[-,postaction={decorate}] (2.4,1.0)--(3.0,1.0); 
\draw[-,postaction={decorate}] (3.2,1.0)--(3.8,1.0); 

\draw[gluon2] (0.3,0.0)--(0.3,0.5); 
\draw[gluon2] (2.3,0.0)--(2.3,0.5); 
\draw[gluon2] (0.6,0.5)--(0.6,1.0); 
\draw[gluon2] (1.4,0.5)--(1.4,1.0); 
\draw[gluon2] (2.6,0.5)--(2.6,1.0); 
\draw[gluon2] (3.4,0.5)--(3.4,1.0);

\node[black] at (-0.1,0.1) {\small \( 1  \)};
\node[black] at (-0.,0.6) {\small \( 2  \)};
\node[black] at (0.3,1.1) {\small \( 3  \)};
\node[black] at (1.1,1.1) {\small \( 4  \)};
\node[black] at (2.,0.6) {\small \( 5  \)};
\node[black] at (2.3,1.1) {\small \( 6  \)};
\node[black] at (3.1,1.1) {\small \( 7  \)};

\end{scope}

\end{tikzpicture}
\begin{tikzpicture}[scale=1.4]

\begin{scope}[thick,decoration={
    markings,
    mark=at position 0.74 with {\arrow{stealth}}}
    ]

\draw[-,postaction={decorate}] (0,0)--(1.5,0); 

\end{scope}
\begin{scope}[thick,decoration={
    markings,
    mark=at position 0.77 with {\arrow{stealth}}}
    ] 

\draw[-,postaction={decorate}] (0.1,0.5)--(1.4,0.5); 
\end{scope}

\begin{scope}[thick,decoration={
    markings,
    mark=at position 0.7 with {\arrow{stealth}}}
    ] 
\draw[-,postaction={decorate}] (0.4,1)--(1.4,1); 
\draw[-,postaction={decorate}] (0.7,1.5)--(1.3,1.5); 

\draw[gluon2] (0.3,0.0)--(0.3,0.5); 
\draw[gluon2] (0.6,0.5)--(0.6,1.0); 
\draw[gluon2] (0.9,1.0)--(0.9,1.5);

\node[black] at (-0.1,0.1) {\small \( 1  \)};
\node[black] at (0.,0.6) {\small \( 2  \)};
\node[black] at (0.3,1.1) {\small \( 3  \)};
\node[black] at (0.6,1.6) {\small \( 4  \)};

\end{scope}

\end{tikzpicture}
\begin{tikzpicture}[scale=1.4]

\begin{scope}[thick,decoration={
    markings,
    mark=at position 0.5 with {\arrow{stealth}}}
    ]

\draw[-,postaction={decorate}] (0,0)--(4,0); 

\draw[-,postaction={decorate}] (0.1,0.5)--(0.9,0.5); 
\draw[-,postaction={decorate}] (1.1,0.5)--(1.9,0.5); 
\draw[-,postaction={decorate}] (2.1,0.5)--(2.9,0.5); 
\draw[-,postaction={decorate}] (3.1,0.5)--(3.9,0.5);

\draw[gluon2] (0.3,0.0)--(0.3,0.5); 
\draw[gluon2] (1.3,0.0)--(1.3,0.5); 
\draw[gluon2] (2.3,0.0)--(2.3,0.5); 
\draw[gluon2] (3.3,0.0)--(3.3,0.5);

\node[black] at (-0.1,0.1) {\small \( 1  \)};
\node[black] at (0.1,0.65) {\small \( 2  \)};
\node[black] at (1.1,0.65) {\small \( 3  \)};
\node[black] at (2.1,0.65) {\small \( 4  \)};
\node[black] at (3.1,0.65) {\small \( 5  \)};

\end{scope}

\end{tikzpicture}
\caption{Example `Mario World' diagrams.}
\label{fig:mwdiags}
\end{figure}
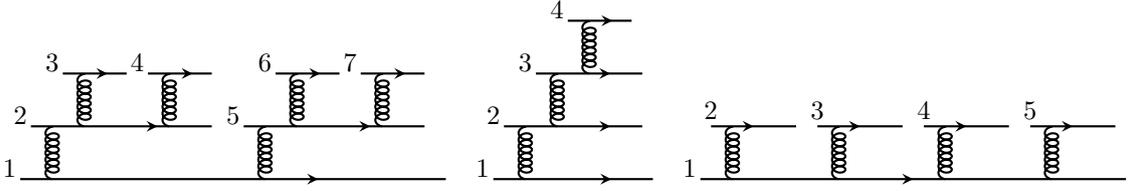

`Mario World' (MW) diagrams are a multi-quark generalisation of the comb diagrams above. There is a one-to-one correspondence between a MW diagram and a permutation $\sig$ in the basis~\eqref{ref:basis}, obtained as follows. Only retain the $s=l$ term in all of the $\Xi_l^a$ operators in the colour factor $C_{1\,\sig\,\ol{1}}$  {\it i.e.} the contribution where the gluon attaches to the highest level quark line; the Feynman diagram which has this colour factor is the  MW Feynman diagram corresponding to the permutation $\sig$. Examples are shown in fig.~\ref{fig:mwdiags} and are drawn with the planar, cyclic ordering $\sig$ that contributes to the corresponding primitive.

Are the MW diagrams UFDs? Some of them are: the subset which have all quark lines at level $l=1$ i.e.
\be
\begin{tikzpicture}[scale=1.4]

\begin{scope}[thick,decoration={
    markings,
    mark=at position 0.5 with {\arrow{stealth}}}
    ]

\draw[-,postaction={decorate}] (0,0)--(4,0); 

\draw[-,postaction={decorate}] (0.1,0.5)--(0.9,0.5); 
\draw[-,postaction={decorate}] (1.1,0.5)--(1.9,0.5); 

\draw[-,postaction={decorate}] (3.1,0.5)--(3.9,0.5); 

\draw[gluon2] (0.3,0.0)--(0.3,0.5); 
\draw[gluon2] (1.3,0.0)--(1.3,0.5); 
\draw[gluon2] (3.3,0.0)--(3.3,0.5); 

\filldraw[fill=black] (2.3,0.3) circle (0.01cm);
\filldraw[fill=black] (2.5,0.3) circle (0.01cm);
\filldraw[fill=black] (2.7,0.3) circle (0.01cm);

\node[black] at (-0.1,0.1) {\small \( 1  \)};
\node[black] at (0.1,0.65) {\small \( a  \)};
\node[black] at (1.1,0.65) {\small \( b  \)};
\node[black] at (3.1,0.65) {\small \( c  \)};

\end{scope}

\end{tikzpicture}
\ee
are unique to the primitives $A(1,a,\ol{a},b,\ol{b},\ldots,c,\ol{c},\ol{1})$, so we have $\mc_{1,a,\ol{a},b,\ol{b},..,c,\ol{c},\ol{1}}=C_{1,a,\ol{a},b,\ol{b},..,c,\ol{c},\ol{1}}$, in agreement with the JO conjecture, eq.~\eqref{eq:jo}. Note that while the flips of the vertices between quark line 1 and the gluons clearly take the primitive out of the basis~\eqref{ref:basis}, so do the $g\,a\,\ol{a}$, $g\,b\,\ol{b}$ $\ldots$ $g\,c\,\ol{c}$ flips, as they change the quark line orientation from the one chosen in \eqref{ref:basis}.

If, however, a quark line nests at level $l>1$, then the vertex where the gluon connects it to the quark line at one lower level {\it can} be flipped and the diagram contribute to a primitive in the basis. For example, for the second MW diagram of fig.~\ref{fig:mwdiags}, there are four such   permutations in the set $\{\text{flip}\}$,
\begin{center}
\begin{tikzpicture}[scale=1.4]

\begin{scope}[thick,decoration={
    markings,
    mark=at position 0.74 with {\arrow{stealth}}}
    ]

\draw[-,postaction={decorate}] (0,0)--(1.5,0); 

\end{scope}
\begin{scope}[thick,decoration={
    markings,
    mark=at position 0.77 with {\arrow{stealth}}}
    ] 

\draw[-,postaction={decorate}] (0.1,0.5)--(1.4,0.5); 
\end{scope}

\begin{scope}[thick,decoration={
    markings,
    mark=at position 0.7 with {\arrow{stealth}}}
    ] 
\draw[-,postaction={decorate}] (0.4,1)--(1.4,1); 
\draw[-,postaction={decorate}] (0.7,1.5)--(1.3,1.5); 

\draw[gluon2] (0.3,0.0)--(0.3,0.5); 
\draw[gluon2] (0.6,0.5)--(0.6,1.0); 
\draw[gluon2] (0.9,1.0)--(0.9,1.5);

\node[black] at (-0.1,0.1) {\small \( 1  \)};
\node[black] at (0.,0.6) {\small \( 2  \)};
\node[black] at (0.3,1.1) {\small \( 3  \)};
\node[black] at (0.6,1.6) {\small \( 4  \)};

\node[black] at (0.7,-0.4) {\small \( 1\,2\,3\,4\,\ol{4}\,\ol{3}\,\ol{2}\,\ol{1}  \)};

\end{scope}

\end{tikzpicture}
\begin{tikzpicture}[scale=1.4]

\begin{scope}[thick,decoration={
    markings,
    mark=at position 0.74 with {\arrow{stealth}}}
    ]

\draw[-,postaction={decorate}] (0,0)--(1.5,0); 
\draw[-,postaction={decorate}] (0.1,0.5)--(1.4,0.5); 
\draw[-,postaction={decorate}] (0.4,1)--(0.9,1); 
\draw[-,postaction={decorate}] (0.9,1)--(1.1,1.5); 
\draw[-,postaction={decorate}] (1.2,1.3)--(1.8,1.1); 

\draw[gluon2] (0.3,0.0)--(0.3,0.5); 
\draw[gluon2] (0.6,0.5)--(0.6,1.0); 
\draw[gluon2] (0.9,1.0)--(1.5,1.2);

\node[black] at (-0.1,0.1) {\small \( 1  \)};
\node[black] at (0.,0.6) {\small \( 2  \)};
\node[black] at (0.3,1.1) {\small \( 3  \)};
\node[black] at (1.8,1.25) {\small \( 4  \)};

\node[black] at (0.7,-0.4) {\small \( 1\,2\, 3\,\ol{3}\,4\,\ol{4}\,\ol{2}\,\ol{1}  \)};

\end{scope}

\end{tikzpicture}
\begin{tikzpicture}[scale=1.4]

\begin{scope}[thick,decoration={
    markings,
    mark=at position 0.74 with {\arrow{stealth}}}
    ]

\draw[-,postaction={decorate}] (0,0)--(1.5,0); 
\draw[-,postaction={decorate}] (0.1,0.5)--(0.6,0.5); 
\draw[-,postaction={decorate}] (0.6,0.5)--(0.9,1.3); 
\draw[-,postaction={decorate}] (0.9,0.9)--(1.8,0.9); 
\draw[-,postaction={decorate}] (1.,1.4)--(1.8,1.4); 

\draw[gluon2] (0.3,0.0)--(0.3,0.5); 
\draw[gluon2] (0.6,0.5)--(1.2,0.9); 
\draw[gluon2] (1.3,0.9)--(1.3,1.4);

\node[black] at (-0.1,0.1) {\small \( 1  \)};
\node[black] at (0.,0.6) {\small \( 2  \)};
\node[black] at (1.9,1.) {\small \( 3  \)};
\node[black] at (1.9,1.5) {\small \( 4  \)};

\node[black] at (0.7,-0.4) {\small \( 1\,2\,\ol{2}\, 3\,4\,\ol{4}\,\ol{3}\,\ol{1}  \)};

\end{scope}

\end{tikzpicture}
\begin{tikzpicture}[scale=1.4]

\begin{scope}[thick,decoration={
    markings,
    mark=at position 0.74 with {\arrow{stealth}}}
    ]

\draw[-,postaction={decorate}] (0,0)--(1.5,0); 
\draw[-,postaction={decorate}] (0.1,0.5)--(0.6,0.5); 
\draw[-,postaction={decorate}] (0.6,0.5)--(0.9,1.3); 
\draw[-,postaction={decorate}] (0.6+0.25,-0.2+1)--(0.6+0.9,-0.2+1); 
\draw[-,postaction={decorate}] (0.6+0.9,-0.2+1)--(0.6+1.1,-0.2+1.5); 
\draw[-,postaction={decorate}] (0.6+1.2,-0.2+1.3)--(0.6+1.8,-0.2+1.1);

\draw[gluon2] (0.3,0.0)--(0.3,0.5); 
\draw[gluon2] (0.6,0.5)--(1.2,0.8); 
\draw[gluon2] (0.6+0.9,-0.2+1)--(0.6+1.5,-0.2+1.2); 

\node[black] at (-0.1,0.1) {\small \( 1  \)};
\node[black] at (0.,0.6) {\small \( 2  \)};
\node[black] at (1.1,0.95) {\small \( 3  \)};
\node[black] at (2.4,1.1) {\small \( 4  \)};

\node[black] at (0.7,-0.4) {\small \( 1\,2\,\ol{2}\, 3\,\ol{3}\,4\,\ol{4}\,\ol{1}  \)};

\end{scope}

\end{tikzpicture}
\end{center}
with relative sign $(+,-,-,+)$. (The first diagram of fig.~\ref{fig:mwdiags} yields sixteen valid flip permutations.)
A crucial observation is that all permutations in $\{\text{flips}\}$ other than the one in correspondence with the MW diagram have a lower value of $l_{\text{tot}}$. This means that if we order the labelling of the $C_i$ and the corresponding $\mc^{MW}_{i}$ such that as $i$ increases, $l_{\text{tot}}$ does not decrease, then eq.~\eqref{eq:matrel} has the form
\be
 \left(   
 \begin{array}{c }
 \\
 \\
\hdashline
\mc^{MW}_1 \\
\mc^{MW}_2 \\
.\\
.\\
.\\
\mc^{MW}_{(n-2)!/k!} \\
\hdashline
\\
\\
\end{array}
\right) 
=
\left(   
\begin{array}{c c c c c c}
 &  & &  &   \\
 &  & &  &   \\
\hdashline
1 & 0 &0  &0  & \ldots & 0 \\
 & 1 & 0 &0   &\ldots  &  0\\
 &  & 1 & 0&  &  . \\
 &  &  & . & &  . \\
 &  &  & & . &  0 \\
 &  & &   &  & 1\\
\hdashline
 &  & &  &   \\
 &  & &  &   \
\end{array}
\right)
 \cdot
 \left(   
 \begin{array}{c }
C_{1} \\
C_{2} \\
.\\
.\\
.\\
C_{(n-2)!/k!} \\
\end{array}
\right)\,,
\ee
and one can obtain a colour factor $C_i$ in terms of its corresponding MW colour factor $\mc^{MW}_i$ and colour factors of primitives with lower $l_{\text{tot}}$. To complete the proof, we show that the colour factor $C_i$ is given by  eq.~\eqref{eq:rec3}, and thus is of the form eq.~\eqref{eq:jo}.

\subsection{Recovering eq.~\eqref{eq:rec3} with MW diagrams}
\label{2p5}

In the previous section, I showed that all the primitives with a unique MW diagram have colour factors  $C_{1,a,\ol{a},b,\ol{b},..,c,\ol{c},\ol{1}}$ which are given by eq.~\eqref{eq:jo}. Let us now assume that all primitive colour factors which have $l_{\text{tot}}\le q-1$ are given by eq.~\eqref{eq:jo}. We aim to show that a colour factor with $l_{\text{tot}}=q$ is also given by eq.~\eqref{eq:jo}.

Let $C_{\sig_q}$ be a colour factor with  $l_{\text{tot}}=q$, and which has a leading quark line $a$, appearing in $\sig_q$ as $..a\,\ol{a}\,\ol{b}..$ (if there is no leading quark line $a$ then we are done, as this is a primitive of the form $C_{1,a,\ol{a},b,\ol{b},..,c,\ol{c},\ol{1}}$). Denote the Feynman diagram colour factor of the  corresponding MW diagram $\mc^{MW}_{\sig_q}$. The relation between the $\mc^{MW}_{\sig_q}$ and primitive colour factors is given by a sum over flip permutations, as described in the previous subsection, and via eq.~\eqref{eq:sigrel}, 
\be
\mc^{MW}_{\sig_q} &=& \sum_{\rho \,\in \,\text{flips}}C_{\rho} \,.
\ee
Let us separate out two of the terms in this sum: the permutation $\sig_q$, and the permutation reached from $\sig_q$ by a single flip of the vertex $g\,b\,\ol{b}$ attaching the leading quark line $a$ to the quark line $b$ one level down---denote this permutation $\sig_q|  \ol{b} \, a \,\ol{a} $. I call the remaining set of permutations $\{\text{flips}'\}$. This set can be further split up into two depending on whether the vertex $g\,b\,\ol{b}$ is flipped or not: denote these sets $\{\text{flips}'|\ol{b} a\ol{a}\}$ and $\{\text{flips}'| a\ol{a}\ol{b}\}$, respectively.  As emphasised in the previous subsection, the primitive colour factors of all permutations other than $\sig_q$ have $l_{\text{tot}}<q$ and so are, by assumption, known through eq.~\eqref{eq:jo}, and therefore satisfy eq.~\eqref{eq:rec3}. We find,
\be
 \mc^{MW}_{\sig_q}&=& C_{\sig_q} -C_{ \sig_q |   \ol{b} \, a \,\ol{a}  } + \sum_{\rho\,\in \,\text{flips}'| a\ol{a}\ol{b}} (-1)^\rho C_{\rho } -  \sum_{\rho\,\in \,\text{flips}'| \ol{b}\, a\ol{a}} (-1)^\rho C_{\rho } \\
  &=& C_{\sig_q}-C_{ \sig_q |    \ol{b} \, a \,\ol{a}  } +  \sum_{\rho\,\in\, \text{flips}'| \ol{ab}} (-1)^\rho C_{\rho }  \\
  &=& C_{\sig_q}-C_{ \sig_q |    \ol{b} \, a \,\ol{a}  } + \mc^{MW}_{\sig_q}- C_{\sig_q|  \ol{ab} } \,.
  \label{eq:final}
\ee
In the second equality above, I have grouped the colour factors with permutations given by $\{\text{flips}'\}$ into pairs $C_{i| a\ol{a}\ol{b}}-C_{i| \ol{b}a\ol{a}}$ and used eq.~\eqref{eq:rec3} to rewrite this as a colour factor with a single particle $\ol{ab}$; the final equality follows from applying eq.~\eqref{eq:sigrel} to the MW colour factor but treating $a \ol{a}\ol{b}$  as a single particle $\ol{ab}$.
Rearranging eq.~\eqref{eq:final}, we find
\be
C_{\sig_q}=C_{\sig_q |  \ol{ab} } + C_{ \sig_q |     \ol{b} \, a \,\ol{a}  } \,,
\ee
that is, that the primitive colour factor is given by the recursion relation eq.~\eqref{eq:rec3}. (In the case where $\{\text{flips}\}$ has only two members---when the leading quark line $a$ is at level $l=2$ and all other quark lines are at level $l=1$---we find $\mc^{MW}_{\sig_q}= C_{\sig_q |  \ol{ab} }$.) This completes the proof.

\acknowledgments
I would like to thank Alexander Ochirov for valuable comments on a draft of this manuscript. This work is supported by U.S. DOE grant DE-AC02-05CH11231. The author additionally acknowledges computational resources provided through ERC grant number 291377: ``LHCtheory''.


\bibliography{./bibliography}
\bibliographystyle{JHEP}

\end{document}